\documentclass[a4paper]{jpconf}
\usepackage{graphicx}
\begin{document}
\title{Azimuthal elliptic anisotropy ($v_{_{2}}$) of high-$p_{_{T}}$ direct $\gamma$ in Au+Au collisions at
$\sqrt{s_{_{NN}}}$ = 200 GeV }
\author{Ahmed M. Hamed (STAR Collaboration)}
\address{University of Mississippi, Oxford, USA\\
Texas A$\&$M University, College Station, USA}
\ead{amhamed@olemiss.edu, ahamed@comp.tamu.edu}

\begin{abstract}
Preliminary results from the STAR collaboration for the azimuthal elliptic anisotropy $(v_{_{2}})$ 
of high transverse momentum ($p_{_{T}}$) direct photons ($\gamma_{_{dir}}$) produced at mid-rapidity ($|\eta^{\gamma_{_{dir}}}|<1$) in Au+Au collisions 
at center-of-mass energy $\sqrt{s_{_{NN}}}=200$~GeV are presented, and compared to the measured $(v_{_{2}})$ of neutral pions ($\pi^{0}$) in the same kinematic range.
The electromagnetic transverse shower profile is used to distinguish $\pi^{0}$ from direct photons. 
The measured $v_{_{2}}^{\gamma_{_{dir}}}(p_{_{T}})$ at high $p_{_{T}}$ ($8< p_{_{T}}^{\gamma_{_{dir}}}<20$~GeV/$c$) is found to be smaller than that of $\pi^{0}$ and consistent 
with zero when using the forward detectors in determining the event plane.  
%Systematic studies are currently in progress.
\end{abstract}
\section{Introduction}
The azimuthal distribution of the produced particles in heavy-ion collisions is expected
to be sensitive to the initial geometric overlap of the colliding nuclei, and would result in
anisotropic azimuthal distributions with respect to the event plane.
The standard method to quantify the azimuthal elliptic anisotropy is to expand the particle azimuthal 
distributions in a Fourier series $\frac{dN}{d\phi} (p_{_{T}})= \frac{N}{2\pi} [ 1 + \sum_{n} 2v_{n}(p_{_{T}})\cos(n(\phi_{p_{_{T}}} - \psi_{_{\textsc{\scriptsize{EP}}}}))]$,
where $\phi_{p_{_{T}}}$ is the azimuthal angle of the produced particle with certain value of 
$p_{_{T}}$, $\psi_{_{\textsc{\scriptsize{EP}}}}$ is the azimuthal angle of the event plane, and $v_{_{n}}$ is the coefficient of the
$n^{th}$ harmonic. The $2^{nd}$ Fourier moment ($n=2$) is referred to as the ``elliptic flow" parameter in the context of the hydrodynamical descriptions, $v_{_{2}}$ and its differential form 
is given by
\begin{equation}
v_{_{2}}(p_{_{T}}) = \langle \langle e^{2i(\phi_{p_{_{T}}}-\psi_{_{\textsc{\scriptsize{EP}}}})} \rangle \rangle = \langle \langle \cos 2(\phi_{p_{_{T}}}-\psi_{_{\textsc{\scriptsize{EP}}}})\rangle \rangle,
\label{eq:TSP}
\end{equation}
where the brackets denote statistical averaging over 
%different 
particles and events.

While RHIC data show large amount of elliptic flow as predicted by the hydrodynamic models at low $p_{_{T}}$, the results
at high $p_{_{T}}$~\cite{STAR_white} are not expected to follow hydrodynamic behavior.  
The medium-induced radiative energy
loss of partons (jet-quenching) has been proposed as the source for the large observed azimuthal elliptic anisotropy at high $p_{_{T}}$,
due to the path-length dependence of the
parton energy loss~\cite{Shuryak}. The STAR results~\cite{STAR1} show the amount of $v_{_{2}}$ at high $p_{_{T}}$ is 
larger than the predicted values 
by pure jet-quenching models. 
Although recent measurements by PHENIX~\cite{PHENIX0} show the produced $\pi^{0}$'s in-plane outnumber those 
produced out-of-plane which may be consistent with the path-length dependence of energy loss, the event plane determination
might have remaining bias toward the direction of the produced jets.
On the other hand, STAR results on the suppression of  direct $\gamma$-triggered vs. $\pi^{0}$-triggered correlated yields~\cite{STAR2} show 
no sensitivity to the path length dependence of parton energy loss. 
If the $v_{_{2}}^{\gamma_{_{dir}}}$ can be measured without bias in the event-plane determination, then the measured value can help 
disentangle the various 
scenarios of direct photon production 
through the expected opposite contributions to the $v_{_{2}}$~\cite{Theory1,Theory2,Theory3,Theory4}, and therefore
could help to confirm the observed binary scaling of the direct photon$~\cite{PHENIX2}$. 
\section{Analysis and Results}
\subsection{Electromagnetic neutral clusters}
The STAR detector is well suited for measuring azimuthal angular correlations 
due to the large coverage in pseudorapidity %($|\eta|<1$)
and full coverage in azimuth ($\phi$). 
While the Barrel Electromagnetic
Calorimeter (BEMC)~\cite{STAR_BEMC} measures the 
electromagnetic energy with high resolution, the Barrel Shower Maximum Detector (BSMD) provides fine spatial 
resolution and enhances the rejection power for the hadrons. The Time Projection Chamber (TPC: $|\eta|<1$)~\cite{STAR_TPC} identifies 
charged particles, measures their momenta, and allows for a charged-particle veto cut with the BEMC matching.   
The Forward Time Projection Chamber (FTPC: $2.4 <|\eta|< 4.0$)~\cite{STAR_FTPC} is used to measure the charged particles momenta and to reconstruct the event plane angle in this analysis.
Using the BEMC to select events (\textit{i.e.} ``trigger") with high-$p_{_{T}}$ $\gamma$,
the STAR experiment collected an integrated 
luminosity of 535~$\mu$b$^{-1}$ of Au+Au 
collisions in 2007 and 973~$\mu$b$^{-1}$ of Au+Au 
collisions in 2011. 
In this analysis, events having primary vertex within $\pm 55$ cm 
of the center of TPC along the beamline, and
with at least one electromagnetic cluster  
with $E_{_{T}} > 8$~GeV are selected. More than 97$\%$
of these clusters have deposited energy greater than 0.5 GeV in each layer
of the BSMD. A trigger tower is rejected if it has a track 
with $p > 3.0 $~GeV/$c$ pointing to it, which reduces the number of the electromagnetic clusters by
only $\sim 7$\%. 
\subsection{$v_{_{2}}$ of neutral and charged particles}
The $v_{_{2}}$ is determined using the standard method (Eq. 1), which 
correlates each particle with the event plane determined from all charged particles with $p_{_{T}} < 2 $~GeV/$c$ (minus
the particle of interest). The event plane is determined by 
\begin{equation}
\psi_{_{\textsc{\scriptsize{EP}}}} = \frac{1}{2} \tan^{-1} (\frac{\sum_{i}\sin(2\phi_{i})}{\sum_{i}\cos(2\phi_{i})} ),
\end{equation}
where $\phi_{i}$ are the azimuthal angles of all the particles used to define the event plane. In this analysis,
the charged-track quality criteria are similar to those 
used in previous STAR analyses~\cite{flow}. The event plane is measured using different techniques and detectors:
%twice: 
1) using all the selected tracks inside the TPC (full-TPC), 2) using the selected tracks in the opposite pseudorapidity side to the particle of interest (off-$\eta$), 
and 3) using all tracks inside the FTPC (full-FTPC)  
in order to reduce the ``non-flow" contributions (azimuthal correlations not related to the event plane). 
Since the event plane is only an approximation to the true reaction plane, 
the observed correlation is divided by the event plane resolution. The event plane resolution is estimated
using the sub-event method in which the full event is 
divided up randomly into two sub-events (for full-TPC, off-$\eta$, and full-FTPC separately) as described in~\cite{flow2}. Biases due to the finite acceptance of 
the detector, which cause the 
particles to be azimuthally anisotropic in the laboratory system are removed 
according to the method in~\cite{shift}.
\subsection{Transverse shower profile analysis}
A crucial part of the analysis is to discriminate between showers from $\gamma_{_{dir}}$ and 
two close $\gamma$'s from high-$p_{_{T}}$ $\pi^{0}$ symmetric decays. 
At $p_{_{T}}^{\pi^0} \sim 8$~GeV/$c$, the angular separation between the two $\gamma$'s 
resulting from a $\pi^{0}$ decay is small, but a $\pi^{0}$ shower is generally broader than a single $\gamma$ shower. 
The BSMD is capable of 
$(2\gamma$)/$(1\gamma)$ separation up to $p_{_{T}}^{\pi^0} \sim 20$~GeV/$c$ due 
to its high granularity ($\Delta\eta\sim 0.007$, $\Delta\phi\sim 0.007$). 
The shower shape is quantified as the cluster energy, 
measured by the BEMC, normalized by the position-weighted energy moment, 
measured by the BSMD strips~\cite{STAR2}.
The shower profile cuts were tuned to obtain a nearly $\gamma_{_{dir}}$-free 
($\pi^{0}_{rich}$) sample and a sample rich in $\gamma_{_{dir}}$ ($\gamma_{_{rich}}$). 
Since the shower-shape analysis 
is only effective for rejecting two close $\gamma$ showers, the $\gamma_{_{rich}}$ sample 
contains a mixture of direct photons and contamination from 
fragmentation photons ($\gamma_{_{frag}}$) and photons from asymmetric hadron ($\pi^0$ and $\eta$) decays.
\subsection{v$_{_{2}}$ of direct photons}
The $\it v_{_{2}}^{\gamma_{_{dir}}}$ is given by:
\begin{equation}
v_{_{2}}^{\gamma_{_{dir}}}=\frac {v_{_{2}}^{\gamma_{_{rich}}}- {\cal{R}}v_{_{2}}^{\pi^{0}}} {1-\cal{R}}, 
\end{equation}
where $\cal{R}$=$\frac{N^{\pi^{0}}}{N^{\gamma_{_{rich}}}}$, and the numbers of
$\pi^{0}$ and $\gamma_{_{rich}}$ triggers
are represented by $N^{\pi^{0}}$ and 
$N^{\gamma_{_{rich}}}$ respectively. 
The value of $\cal{R}$ is measured in~\cite{STAR2} and 
found to be 
$\sim 30\%$ in central Au+Au. 
In Eq. 3 all background sources for $\gamma_{_{dir}}$ are assumed to have the same $v_{_{2}}$ as $\pi^{0}$. 
Thus, all remaining background is subtracted under the assumption that the background particles have the same correlation functions as that measured for 
$\pi^0$ triggers.
\begin{figure}
\begin{center}
   \resizebox{100mm}{150pt}{\includegraphics{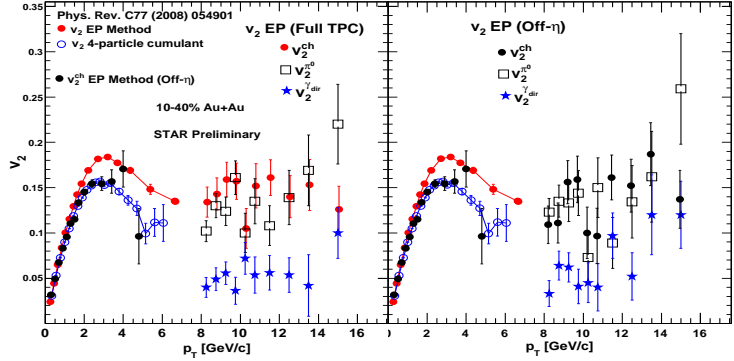}} 
        \caption{(Color online) For $p_{_{T}} < 6$~GeV/$c$, (Au+Au 2007) both panels show previous STAR measurements~\cite{flow} of $v_{_{2}}$
	as a function of $p_{_{T}}$ for charged particles with $|\eta| <$ 1 in 10-40\% $Au+Au$
	collisions at $\sqrt{s_{_{NN}}}=200$~GeV using the Event-Plane method (closed red circles), and the 4-particle cumulant
	method (open circles). Also $v_{_{2}}$ for charged particles ($|\eta| <$ 1) using off-$\eta$ event plane method is shown in closed black
	circles (this analysis). For $p_{_{T}} > 6$~GeV/$c$: (Au+Au 2007)
	both panels show $v_{_{2}}$ of
	charged particles, $\pi^{0}$, and $\gamma_{_{dir}}$ (circles, squares, stars respectively) using the full TPC (left
	panel) and using the off-$\eta$ method (right panel). 
	}
    \label{fig:corrfunc}
    \end{center}
\end{figure} 
\begin{figure}
\begin{center}
   \resizebox{100mm}{150pt}{\includegraphics{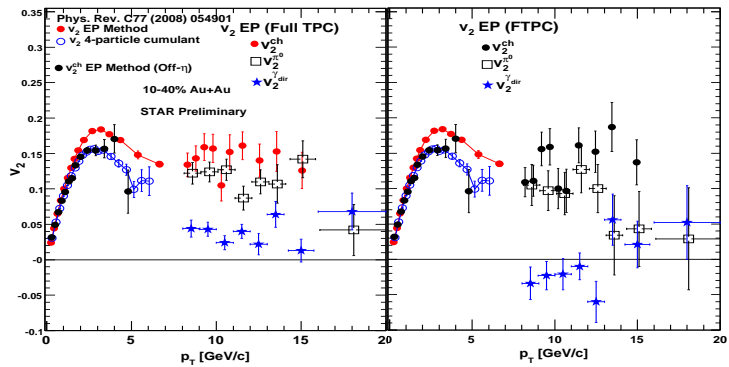}} 
        \caption{
	(Color online) For $p_{_{T}} < 6$~GeV/$c$, (Au+Au 2007) both panels show measurements as in Fig. 1.
        For $p_{_{T}} > 6$~GeV/$c$, (Au+Au 2011 for $\pi^{0}$ and $\gamma_{_{dir}}$)
	both panels show $v_{_{2}}$ of
	charged particles, $\pi^{0}$, and $\gamma_{_{dir}}$ (circles, squares, stars respectively) using the full TPC (left
	panel) and using full FTPC (right panel). 
	}
    \label{fig:corrfunc}
     \end{center}
\end{figure} 
Figure 1 (both panels) shows the $v_{_{2}}$ of  
charged particles ($v_{_{2}}^{ch}$) at low $p_{_{T}}$ ($p_{_{T}} < 6$~GeV/$c$) using the event plane method (off-$\eta$)  
compared to previous STAR measurements~\cite{flow}, and 
the $v_{_{2}}$ of the charged particles, neutral pions and direct photons using the full-TPC and off-$\eta$ event plane methods at high $p_{_{T}}$ ($p_{_{T}} > 6$~GeV/$c$). 
At low $p_{_{T}}$ the $v_{_{2}}^{ch}$(off-$\eta$) is smaller than the $v_{_{2}}$ using the full TPC and agrees well with 
the $v_{_{2}}$$\{4\}$ (4-particle cumulant) method, in which 
the contribution of the non-flow is expected to be small. 
At high $p_{_{T}}$ the two different
methods (full TPC and off-$\eta$) for the event plane measurements give similar results 
, which might 
indicate that the off-$\eta$ method is not free from a bias in the event-plane determination.
While the $v_{_{2}}^{\pi^{0}}$ and $v_{_{2}}^{ch}$ are similar ($\sim$ 12$\%$), the $v_{_{2}}^{\gamma_{_{dir}}}$ 
is systematically lower than that of hadrons. 
The similarity of the $v_{_{2}}$ results using the full-TPC and off-$\eta$ at high $p_{_{T}}$, along with 
the non-zero value of $v_{_{2}}^{\gamma_{_{dir}}}$, indicate a remaining bias in the event-plane determination.

Figure 2 (left and right panels) shows the $v_{_{2}}^{\pi^{0}}$ and $v_{_{2}}^{\gamma_{_{dir}}}$ for ($8< p_{_{T}}^{\gamma_{_{dir}}}<20$~GeV/$c$) 
from Au+Au 2011 data using the full-TPC ($|\eta|<1$) and full-FTPC ($2.4 <|\eta|< 4.0$). 
The 
results from two different data sets (Au+Au 2007 and Au+Au 2011) are consistent (left panels of Fig. 1 and Fig. 2) using the full-TPC. 
While using the FTPC in determining the event plane (Fig. 2 - right panel) the $v_{_{2}}^{\gamma_{_{dir}}}$ is consistent with zero. Assuming the dominant source of direct photons is
prompt hard production, the zero value implies no remaining bias in the event-plane determination. Accordingly, the measured value of 
$v_{_{2}}^{\pi^{0}}$ would be the effect of path-length dependent energy loss. Systematic studies are currently in progress.
\section{Conclusions}
The STAR experiment has reported the first $v_{_{2}}^{\gamma_{_{dir}}}$ at high-$p_{_{T}}$ ($8< p_{_{T}}^{\gamma_{_{_{dir}}}}<20$~GeV/$c$) at RHIC. 
Using the mid-rapidity detectors in determining the event plane, the measured value of $v_{_{2}}^{\gamma_{_{dir}}}$ is non-zero, and is probably due to biases in the event-plane determination. 
Using the forward detectors in determining the event plane could eliminate remaining biases, and the measured $v_{_{2}}^{\gamma_{_{dir}}}$ is consistent with zero.
The zero value of 
$v_{_{2}}^{\gamma_{_{dir}}}$ suggests a negligible contribution of jet-medium photons~\cite{Theory2}, and negligible effects 
of ${\gamma_{_{frag}}}$~\cite{Theory1} on the $v_{_{2}}^{\gamma_{_{dir}}}$ over the covered kinematics range. The measured value of 
$v_{_{2}}^{\pi^{0}}$, using the forward detectors in determining the event plane, is apparently due to the path-length dependence of energy loss.

\end{document}